\documentclass[twocolumn,showpacs,preprintnumbers,amsmath,amssymb]{revtex4}

\def\be{\begin{equation}}
\def\ee{\end{equation}}
\def\bee{\begin{eqnarray}}
\def\eee{\end{eqnarray}}

\input{epsf}
\newcommand{\epsfig}[3]{%
    \begin{figure}[htbp]%
        \begin{center}%
            \setlength{\unitlength}{1.0cm}%
            \begin{picture}(7,9)%
                \epsfxsize=7cm%
                \epsffile{#1}%
                \end{picture}%
            \end{center}%
        \caption{#3}%
        \label{#2}%
        \end{figure}%
 }

\begin{document}
\title{On the extrapolation to ITER of discharges in pressent tokamaks}

\author{A.G. Peeters, C. Angioni, A.C.C. Sips, and the ASDEX Upgrade Team}

\address{Max Planck Institut f\" ur Plasmaphysik, Botzmannstrasse 2 \\
D-85748 Garching bei M\" unchen, Germany}

\begin{abstract}
An expression for the extrapolated fusion gain $G = P_{\rm fusion}/5 P_{\rm heat}$
($P_{\rm fusion}$ being the total fusion power and $P_{\rm heat}$ the total heating 
power) of ITER in terms of the confinement improvement factor ($H$) and the 
normalised beta ($\beta_N$) is derived in this paper.
It is shown that an increase in normalised beta can be expected to have a 
negative or neutral influence on $G$ depending on the chosen confinement
scaling law. 
Figures of merit like $H \beta_N/ q_{95}^2$ should be used with care, since 
large values of this quantity do not guarantee high values of $G$, 
and might not be attainable with the heating power installed on ITER. 
\end{abstract}


\maketitle

\section{INTRODUCTION}

In many tokamak devices discharge scenarios are studied with
the aim of improving the performance of future reactor experiments
over the current design values. 
Essentially two ingredients enter in the optimisation: the energy 
confinement time and the Magneto-Hydro-Dynamic (MHD) stability limit,
represented by a critical pressure. 
Both energy confinement time and obtainable pressure are measured in 
current experiments and then for scaling purposes expressed in dimensionless
parameters. 
The confinement improvement is given by the so-called $H$-factor which 
measures the energy confinement time ($\tau_E$) relative to a scaling law
($\tau_{E,scaling}$). 
The obtainable volume averaged pressure ($\langle p \rangle$) is expressed 
in the normalised quantity $\beta_N$
\be
H = \tau_{\rm E} / \tau_{\rm E,scaling} \quad
\beta_{\rm N} = \langle \beta \rangle { a B \over I_p}.
\label{definitions}
\ee  
In the equations above $\langle \beta \rangle$ is the volume average of 
$2 \mu_0 p / B^2$ measured in \%, $a$ is the minor radius in m, $B$ is the 
magnetic field in T, and $I_p$ is the plasma current in MA.
For the extrapolation one assumes a constant $H$-factor and then uses the 
confinement scaling to determine the confinement time in a next step device. 
Furthermore, it is assumed that the working point of the reactor is at the
same normalised pressure $\beta_N$. 
The $H$-factors describe our imperfect knowledge of the scaling of confinement, 
i.e. the confinement in current day experiments in some areas of the parameter 
space scales differently than the developed scaling laws suggest. 
Of course, the use of a constant $H$-factor is a daring approach to correct for 
this lack of knowledge, but nevertheless can give some idea of how confinement 
could be different in the next step experiment. 

Different scenarios for improved performance have been proposed (see for an 
overview \cite{LIT04}). 
These scenarios include the internal transport barriers (ITB) as well as 
the different scenarios for the improvement of the H-mode. 
The results presented in this paper can in principle be applied to all scenarios,
but as examples only the H-mode scenarios will be shown
The latter scenarios are an active area of research with many contributions 
from different machines, for instance, ASDEX Upgrade \cite{GRU99b}, DIII-D 
\cite{LUC01}, JET \cite{SIP03}, and JT60-U \cite{KAM99}. 
(Although the original reference \cite{GRU99b} refers to an internal transport
barrier it was shown in a later paper \cite{PEE02} that the ion temperature profiles 
follow the same scaling as those of the standard H-mode). 

An improvement in confinement or MHD stability would allow to operate the 
next step tokamak experiment ITER \cite{ITE99,SHI05} at a higher energy 
multiplication factor $Q$, a higher fusion power, or a higher bootstrap current 
fraction. We define the energy multiplication factor $Q$ and the fusion gain $G$ 
as
\be 
Q = {P_{\rm fusion} \over P_{\rm aux}},\qquad
G = {P_{\rm fusion} \over 5 P_{\rm heat}}
\label{qdef}
\ee
where $P_{\rm fusion}$ is the total fusion power, $P_{\rm aux}$ is 
the externally applied heating power, and $P_{\rm heat}$ is the total heating
power of the plasma, which in steady state is equal to the loss power $P_{\rm loss}$.
Since one fifth of the fusion power heats the plasma, $P_{\rm heat} = P_{\rm fusion} / 5 + 
P_{\rm aux}$ and there is a direct relation between $G$ and $Q$
\be 
G = {Q \over Q + 5} \label{qrelation}
\ee
The currently proposed next-step experiment ITER \cite{SHI05} is 
designed to reach $Q=10$, and any realistic scenario to be tested in
this experiment should reach a $Q$ value significantly larger than one. Scenarios at 
sufficiently large $Q$ might be further optimised to reach a higher fusion
power, or to reach a higher bootstrap current fraction. The latter 
optimisation is known as the hybrid scenario. It aims at the extention
of the pulse length at similar performance as the design value. 
It is important for the results presented in this paper to stress that
improvements in fusion power or bootstrap fraction can only be of interest
if the the energy multiplication factor is sufficiently high.
In the opinion of the authors a good representation of the extrapolated 
performance of current discharges towards ITER can only be obtained if one 
of the figures of merit is directly connected with $Q$. Other figures 
could be used to measure the bootstrap current and fusion power. 
Although $Q$ plays a central role in the ITER experiment their is no 
published simple scaling that allows a direct assessment of the extrapolated
value of current discharges. A simple way to obtain a rough estimate can
be extremely useful in assessing the progress made in this large area 
of research. 
Finally, we would like to stress that our extrapolation formula aims at 
judging the performance in ITER and is not necessarily applicable for more
general purposes. 
In particular we use a fixed size and magnetic field. 
A reactor design can also be optimised through changes in these parameters. 

\section{SCALING OF G}

The fusion power is proportional to 
\be 
P_{\rm fusion} \propto  {\langle \sigma v \rangle \over T^2}
V  \,\,p^2,
\label{fusionpower}
\ee
where $T$ is the plasma temperature, $V$ is the plasma volume, and 
$\langle \sigma v \rangle$ is the cross section for the fusion reactions 
averaged over the velocity distribution. 
Over the temperature range of interest $\langle \sigma v \rangle \propto 
T^2$, such that the fusion power scales with the pressure ($p$) squared. 
The power loss from the plasma ($P_{\rm loss}$) is measured by the energy 
confinement time $\tau_E$, and under stationary conditions it is balanced 
by the total heating power 
\be 
P_{\rm loss} = {W \over \tau_E} = {3 p V \over 2 \tau_E} = P_{\rm heat}, 
\label{ploss}
\ee
where $W$ is the stored energy. 
Combining the Eqs~(\ref{qdef}), (\ref{fusionpower}), (\ref{ploss}), and 
(\ref{qrelation}) one obtains 
\be
G = {Q \over Q + 5} =  {P_{\rm fusion} \over 5 P_{\rm heat}} \propto n T \tau_E,
\label{nttaueq}
\ee
where $p = nT$ was used, and $n$ is the plasma density. This is of course 
the famous $n T \tau_E$ product. 

To proceed we write the confinement time as a product of the improvement 
factor $H_X$ over the confinement time of an arbitrary scaling law ($\tau_{X}$)
\be
\tau_E = H_X \tau_{X} = H_X C_X\, I_p^{\alpha_I} \, B^{\alpha_B} \, P_{\rm loss}^{\alpha_P} \, 
n^{\alpha_n} \, M^{\alpha_M} \, a^{\alpha_a} \, \kappa^{\alpha_k}.
\label{tau}
\ee
Here $C_X$ is a constant, $M$ is the effective mass in AMU, and $\kappa$ is 
the plasma elongation. The exponents $\alpha$ are the exponents of the scaling
law. 
For the projection to ITER the plasma size ($a$,$R$,$\kappa$) 
as well as the magnetic field $B$ and the effective mass $M$ are assumed to be 
given by the design values.
For the density a fixed ratio of the Greenwald limit ($n_{Gr}$) \cite{GRE88} will 
be used, i.e. $n \propto n_{Gr} I_p$. In our final result it will not be difficult 
to obtain the result at constant density (i.e. the design value of the density 
without considering a scaling of this density with plasma current) since the density 
dependence that enters can be 
easily identified through the coefficient $\alpha_n$, which can be set 
to zero to obtain the result for a given fixed density. 
Note that we cannot assume $P_{\rm loss}$ to be constant, since discharges 
at different beta will extrapolate to a different fusion power and, hence, 
to different plasma heating powers. 
Since for fixed magnetic field and plasma shape $I_p \propto q_{95}^{-1}$, with $q_{95}$ 
being the safety factor at 95\%\space of the plasma radius, and 
redefining the constant $C_X$ to include all constant design quantities one obtains 
\be
\tau_E = H_X \tau_{X} = H_X C_X\, q_{95}^{-\alpha_I-\alpha_n} \, P_{\rm heat}^{\alpha_P} \, 
\label{taured}
\ee
Combining Eq.~(\ref{taured}) with Eq.~(\ref{nttaueq}) one obtains an expression 
for $G$. Indicating all quantities of the ITER standard scenario with an index $S$, 
one can build the ratio 
\be 
{G \over G_{S}} = {H_X \over H_{XS}} {q_{95S}^{\alpha_I +\alpha_n} \over
q_{95}^{\alpha_I +\alpha_n}} {P_{\rm heat}^{\alpha_P} \over  P_{\rm heat,S}^{\alpha_P}}
{p \over p_S}
\ee
Then recalling that $P_{\rm heat} = P_{\rm fusion}/5G$ and $P_{\rm fusion} \propto p^2 
\propto (\beta_N/q_{95})^2$ one arrives at 
\be 
{G \over G_S} = {H_X \over H_{XS}} 
\biggl ({G \over G_S}\biggr )^{-\alpha_p}
\biggl ( {q_{95S} \over q_{95}} \biggr )^{1+2\alpha_p + \alpha_I + \alpha_n} 
\biggl ({\beta_N \over \beta_{NS}} \biggr )^{1+ 2\alpha_p}  
\ee
Combining the terms containing $G$ we finally derive at the desired expression
\be 
{G \over G_S} = \biggl ( {H_X \over H_{XS}}\biggr )^{{1 \over 1+\alpha_p}} 
\biggl ( {q_{95S} \over q_{95}} \biggr )^{{1+2\alpha_p + \alpha_I + \alpha_n \over 1 + \alpha_p}} 
\biggl ({\beta_N \over \beta_{NS}} \biggr )^{{1+ 2\alpha_p \over 1 + \alpha_p}}  
\label{final}
\ee
The figures of merit for different scalings can now be derived directly.
Here the explicit expressions are given for four different scaling lows.  
The most commonly used IPB98(y,2) \cite{ITE99} indicated by $\tau_E^H$, the 
L-mode scaling ITER89-P denoted by $\tau_E^L$, a newly derived scaling from 
Ref.~\cite{COR05} denoted by $\tau_E^C$, and an electro-static gyro-Bohm
scaling law derived in \cite{PET03} denoted by $\tau_E^{EGB}$
\be 
\tau_E^H =  0.145 H_H {I_p}^{0.93}  B^{0.15}  P^{-0.69}  n^{0.41}
M^{0.19} R^{1.39} a^{0.58}  \kappa^{0.78}
\label{98y2}
\ee
\be
\tau_E^L = 0.048 H_L I_p^{0.85} B^{0.2} P_{\rm heat}^{-0.5} n^{0.1}  
M^{0.5} R^{1.2} a^{0.3} \kappa^{0.5} 
\label{89P}
\ee
\be
\tau_E^C  = 0.092 H_C I_p^{0.85} B^{0.17} P_{\rm heat}^{-0.45} n^{0.26} M^{0.11} 
R^{1.21} a^{0.39} \kappa^{0.82}.
\label{cordey}
\ee
\be
\tau_E^{EGB}  = 0.0865 H_{EGB} I_p^{0.83} B^{0.07} P_{\rm heat}^{-0.55} n^{0.49} M^{0.14} 
R^{1.81} a^{0.30} \kappa^{0.75}.
\label{egb}
\ee
In the equations above $n$ is the density in units of $10^{20}$ ${\rm m}^{-3}$,
$R$ is the major radius in m, $a$ is the minor 
radius in m, $M$ is the averaged ion mass in AMU, $\kappa = A / {\pi a^2}$ 
is the plasma elongation, $A$ is the area of the poloidal cross section, and $P_{\rm heat}$
is in MW. 
The new scaling laws ($\tau_E^C,\tau_E^{EGB}$) have been obtained after designed experiments 
have shown a small and possibly absent $\beta$ dependence of the confinement 
\cite{PET03,COR96,PET98,DON04,PET04} in contrast with the IPB98(y,2) scaling which, when 
expressed in normalised quantities (normalised pressure $\beta\propto n T B^{-2}$, normalised 
Larmor radius $\rho_* \propto \sqrt{T}/Ba$, and normalised collisionality 
$\nu_* \propto  n a T^{-2}$) has an unfavourable beta \cite{ITE99} dependence
\be 
B \tau_E \propto \rho_*^{X} \beta^{Y} \nu_*^{Z} \propto 
\rho_*^{-2.7} \beta^{-0.9} \nu_*^{-0.01}. 
\label{dimensionless98}
\ee
For the electro-static gyro-Bohm scaling, therefore, zero beta dependence as well as 
$\tau_E \propto \rho_*^{-3}$ were imposed to derive
\be 
B \tau_E \propto \rho_*^{-3} \beta^0 \nu_*^{-0.14} .
\label{dimensionlessegb}
\ee
Several papers have pointed out the fact that the absence of the beta dependence 
in $\tau_E$ leads at high normalised pressure to more optimistic projections for ITER 
compared with IPB98(y,2) scaling \cite{PET04,COR05}. 

To obtain the scaling expressions for $G$ a standard scenario must be defined.
Here $q_{95S} = 3$ and $\beta_{NS} = 1.8$ will be used. 
The H-factors ($H_{XS}$) can be calculated by dividing the target confinement time by the 
confinement times of the scaling calculated using Eqs.~(\ref{98y2}), (\ref{89P}), 
(\ref{cordey}), and (\ref{egb}). 
In the latter equations the ITER parameters ($I_p = 15$ MA, $B = 5.3$ T, $R = 6.2$ m, 
$\kappa = 1.75$, $n = 10^{20}$ ${\rm m}^{-3}$, $a = 2$ m, $M = 2.5$, $P = 87$ MW, 
$\tau_E = 3.68$ s) are used, yielding $H_{HS} = 1.$, $H_{LS} = 2.2$, $H_{CS} = 1.07$, and 
$H_{EGB,S} = 0.8$. One then directly finds 
\be 
G = C H^X q_{95}^Y \beta_N^Z.
\label{arbit}
\ee
with the values of the constant $C$ and the scaling potential giving in Table~\ref{table}.
From this table it can be seen that $\beta_N$ has a strongly negative effect in the IPB98(y,2)
scaling, and a rather small effect in all the other scalings. It is clear from Eq.~(\ref{final})
that $G \propto \beta_N^0$ occurs only for the generic scaling $\tau_E \propto P^{-0.5}$. 

\begin{table}
\caption{Values of the constants of Eq.~(\ref{arbit}) for the different scalings. 
Both coefficients assuming ITER operation at fixed absolute density as well as 
at fixed Greenwald fraction ($n = 0.85 n_{Gr}$) are given. The latter are indicated
by the letters ``Gr''.\label{table}}
\begin{tabular}{|c||cccc|}
\hline
 & \hbox to 1 truecm{\hfil C \hfil} & \hbox to 1 truecm {\hfil X \hfil} & 
   \hbox to 1 truecm{\hfil Y \hfil } & \hbox to 1 truecm {\hfil Z \hfil}\\
\hline
\hline
IPB98(y,2) & 9.62 & 3.22 & -1.77 & -1.23\\
\hline
IPB98(y,2) Gr & 41.15 & 3.22 & -3.19 & -1.23\\
\hline
ITER89-P & 0.892 & 2.0 & -1.70 & 0 \\
\hline
ITER89-P Gr & 1.11 & 2.0 & -1.90 & 0 \\
\hline
Cordey & 3.53 & 1.82 & -1.72 & 0.18 \\
\hline
Cordey Gr & 5.94 & 1.82 & -2.20 & 0.18 \\
\hline
EGB & 7.41 & 2.22 & -1.62 & -0.22 \\
\hline
EGB Gr & 24.52 & 2.22 & -2.71 & -0.22 \\
\hline
\end{tabular} 
\end{table}

For the derivation of a figure of merit one often approximates the coefficients of 
the scaling law. 
For a better comparison we can make the similar approximations, i.e. $\alpha_I = 1$, 
$\alpha_n = 0$, and $\alpha_p = -2/3$ for the IPB98(y,2) and $\alpha_p = -1/2$ for all 
other scaling laws.
This yields 
\be
G = 10.8 {H_H^3 \over \beta_N q_{95}^2}, 
\label{HHscal}
\ee
and 
\be
G = \gamma_X {H_X^2 \over q_{95}^2} ,
\label{scalzerob}
\ee
with 
\be 
\gamma_L = 1.24 \quad 
\gamma_C = 5.25 \quad
\gamma_{EGB} = 9.375.
\ee
The figure of merit $H \beta_N / q_{95}^2$ can be considered a scaling for $G$ 
if the exponents in the scaling law for the confinement are $\alpha_I = 1$, 
$\alpha_n = 0$, and $\alpha_p = 0$, i.e. only in the ideal condition if it is 
assumed that there is no power degradation. 
We note that one can define a figure of merit in different ways and that 
$H \beta_N / q_{95}^2$ can be thought of as a combination of good confinement, and 
high fusion power \cite{WOL03}. Also for the generic scaling ($\alpha_P = -0.5$)
the bootstrap current in a reactor close to ignition scales as $H \beta_N$ \cite{PEE00}.
It is, however, clear that $H \beta_N / q_{95}^2$ does not provide a figure of 
merrit for $G$. This point is important because it shows that one 
must be careful in using $H \beta_N / q_{95}^2$. 
From the scaling Eq.~(\ref{HHscal}) it follows that the difference with the 
figure of merit $H \beta_N / q_{95}^2$ is $\propto (1.8 H_H / \beta_N)^2$.
The latter quantity is for discharges with normal confinement $H_H = 1$ but 
high normalised pressure $\beta_N = 3.6$ as small as 0.25. 
This makes a large difference in $G$.
The above example with $H_H = 1$, $\beta_N = 3.6$, and $q_{95} = 4.2$ reaches 
a value for the parameter $H_H \beta_N / q_{95}^2$ that suggest that the ITER 
target ($Q = 10$) could be reached, whereas in reality such a discharges would 
extrapolate to $Q = 1$. 
Not only is this value of little interest, it also requires a rather
large heating power, since the fusion power is four times larger and energy multiplication
is much smaller. 
The other scaling laws suffer from the same problem, although it is less dramatic
due to the different exponents of $H$ as well as $\beta_N$.

\epsfig{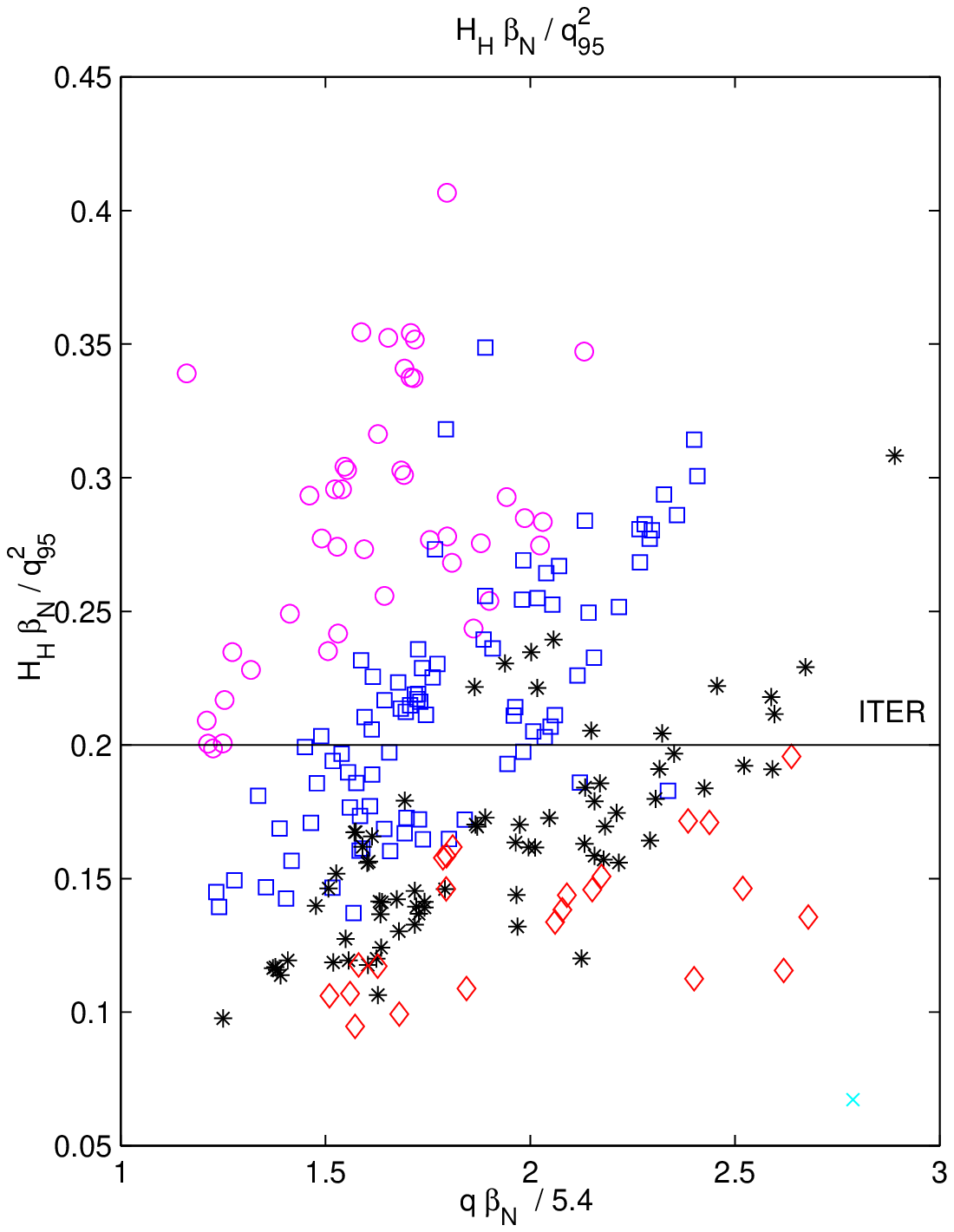}{fig_adv_old}{(color online) Figure of merit $H_H \beta_N / q_{95}^2$
of the advanced scenario discharges from ASDEX Upgrade. The symbols correspond to 
different values of $q$: circles (magneta) $q<3.5$, squares (blue) 
$3.5<q<4.0$, stars (black) $4.0< q<4.5$, diamonds red $4.5 < q < 5.0$, 
crosses $q > 5.0$}

\section{DIMENSIONLESS VARIABLES} 

Some confusion can arise when considering the scaling in terms of the dimensionless
parameters $\beta$, normalised Larmor radius $\rho^* = \rho / a$ and normalised collisonality
$\nu^*$. 
The scaling of $G$ in terms of dimensionless parameters yields \cite{PET04}
\be 
G \propto \beta (B \tau_E) B \propto  \rho_*^{-1.5+X}
\beta^{1.25 + Y} \nu_*^{-0.25 + Z} R^{-1.25}
\label{petty}
\ee
where $X,Y,Z$ are the coefficient of the $\rho_*$, $\beta$ and $\nu_*$ scaling of confinement
as defined in Eq.~(\ref{dimensionless98}). 
In the equation above the dependence on $\rho_*$ as well as $\nu_*$ has been explicitly added 
compared with Ref. \cite{PET04}. 
Using the scaling $G \propto \beta^{1.25+Y}$ it was concluded \cite{PET04} that for 
the IPB98(y,2) scaling ($X = -0.9$) there is no large benefit of going to high $\beta$ since 
$Q$ increases only moderately with $\beta$ ($\beta^{0.35}$), whereas for the energy 
confinement scalings that have no beta dependence it would be largely advantageous to go to 
the $\beta$ limit since $G \propto \beta^{1.25}$. This conclusion seems in disagreement with 
the results derived in this paper, which rather point at a decreasing $G$ with $\beta_N$ for 
the IPB98(y,2) scaling and a $G$ independent of $\beta_N$ for the other scaling laws. 
Eq.~(\ref{petty}) is the correct dimensionless expression, but it must be noted 
that the scaling with beta
$G \propto \beta^{1.25 + Y}$ holds only at constant normalised Larmor radius 
($\rho_*$) as well as collisionality ($\nu_*$). 
The difference with the results in this paper is that the results are evaluated
for a fixed machine size, density, and magnetic field. 
With these assumptions it is not possible to change $\beta$ independently 
of $\rho_*$ and $\nu_*$. 
At fixed density the $\beta$ scaling is essentially a temperature scaling, leading to changes 
in the normalised Larmor radius as well as the collisionality.  
Using the scalings of $\rho_*$ and $\nu_*$ one can derive from Eqs.~(\ref{dimensionless98}),
(\ref{dimensionlessegb}), and (\ref{petty})
\be
G_H \propto T^{-1.23} \qquad
G_{EGB} \propto T^{-0.22}.
\ee
These scalings are consistent with the diagrams of Ref. \cite{PET04} where $G$ can 
be seen to decrease with increasing $T$ even for the electro-static gyro-Bohm scaling.
Therefore we arrive at the conclusion that for a given design reaching the beta 
limit does not help in increasing $G$.  
Of course the density scaling is more hidden in our approach since it is considered
to be a design value. One can derive that 
$G$ increases strongly with density for all scalings.

Finally, it is noted here that for constant heating power $\beta_N \propto H$ and all 
figures of merit have the same form
\be 
{H \beta_N \over q_{95}^2} \rightarrow {H^3 \over \beta_N q_{95}^2} \rightarrow
{H^2 \over q_{95}^2}
\ee

\section{PERFORMANCE DIAGRAM}

Having derived a simple expression that directly allows to evaluate $G$, a suggestion 
is made in this section for a diagram that should allow for an easy assessment of the 
extrapolated performance of any discharge. 
From the discussion above it is clear that a better representation of 
the data can be obtained by plotting the scaling for $G = Q / (Q + 5)$.
against either the scaling of the fusion power, i.e. $\beta_N^2 / q_{95}^2$,
or the scaling of the bootstrap current, i.e. $q_{95} \beta_N$. Since both
are of importance it is useful to mark the different $q_{95}$ values
by different symbols in whatever plot one chooses. This makes that all
important parameters can be estimated from the same graph. 

\epsfig{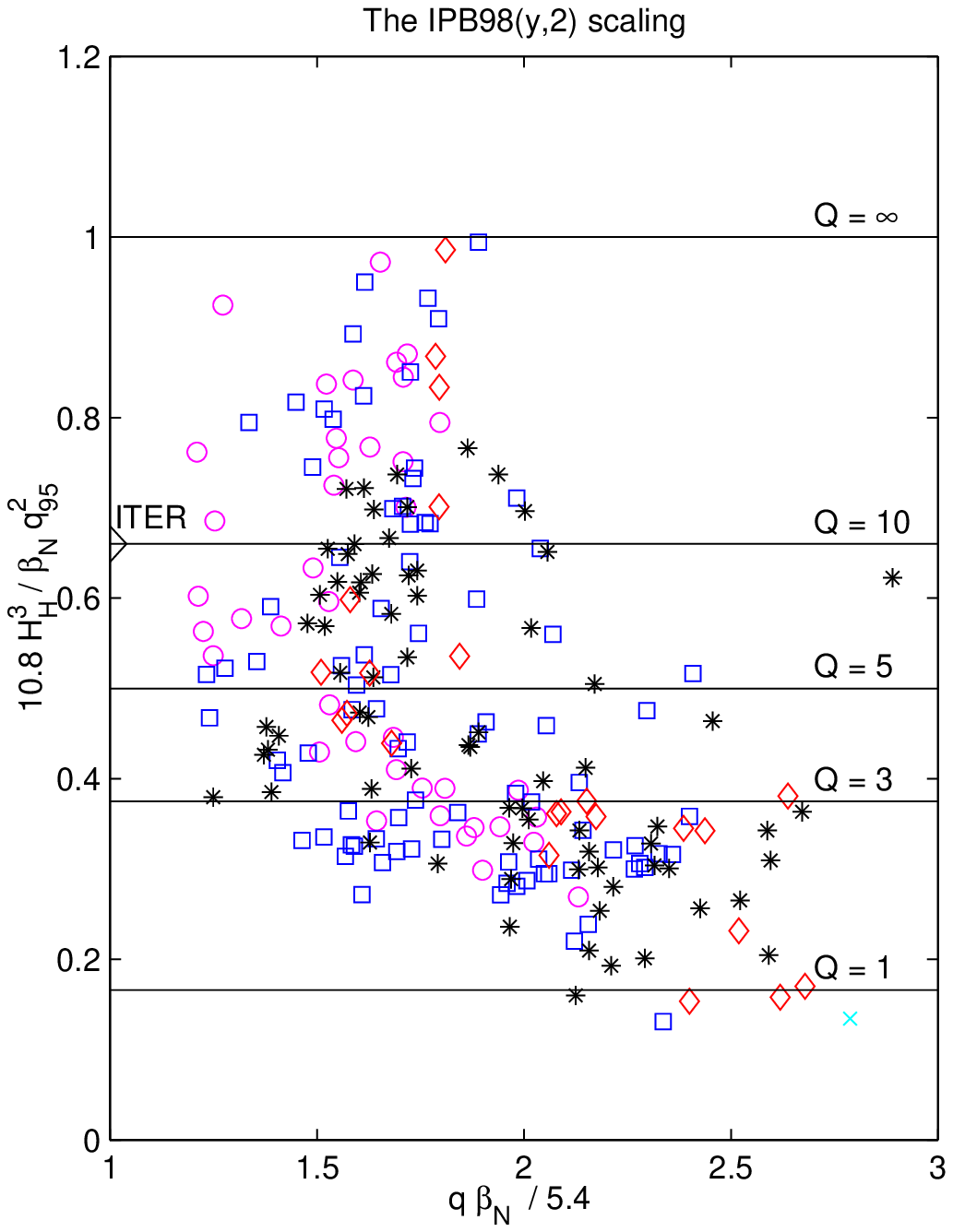}{fig_adv98}{Fig. \ref{fig_adv98} (color online) Scaling for 
the advanced scenario discharges based on the IPB98(y,2) scaling. Symbols reflect 
the $q_{95}$ values as in figure \ref{fig_adv_old}.}

Figure \ref{fig_adv_old} shows a dataset of advanced scenario discharges from 
ASDEX Upgrade 
in the representation using $H_H \beta_N / q_{95}^2$ versus $q_{95} \beta_N$. 
In this figure the different $q_{95}$ values are indicated with with different symbols 
(and colours in the online version). 
In the representation using $H \beta_N / q_{95}^2$ even the points 
at highest $q_{95} \beta_N$ reach the ITER target. 

Figure \ref{fig_adv98} shows the scaling derived from Eq.~(\ref{HHscal}). 
This can be directly compared with Fig.~\ref{fig_adv_old} since the same scaling 
law is used. The obtained picture is different in the sense that the highest 
$q_{95} \beta_N$ values no longer reach the ITER target for $Q$. These discharges 
have only moderate confinement improvements and high $\beta_N$ leading to a relatively 
small $(H_H/ \beta_N)^2$. 
In the diagrams $Q$ can exceed infinity, which is obviously unphysical. For those
discharges for which $Q>\infty$, the temperature will rise until the fusion cross
section no longer scales quadratically with $T$, violating the original assumptions
in the derivation, and leading to a smaller increase of the fusion power with $T$. 
This stabilises the solution and leads to $Q = \infty$. 

In the diagrams presented so far the external heating power is still implicit. 
A better insight of how much external heating power is needed to run a 
certain discharge under reactor conditions can be obtained from the diagram that has 
the scaling of the fusion power $P_{\rm fusion} \propto (\beta_N / q_{95})^2$ on the 
x-axis. Because $P_{\rm aux} = P_{\rm fusion} / Q$ one obtains 
\be 
P_{\rm aux} \propto {1\over Q} \biggl ( {\beta_N \over q_{95}} \biggr )^2. 
\ee
For a fixed auxiliary heating power the relation above determines a curve in the 
$G = Q/(Q+5)$ versus $(\beta_N/q_{95})^2$ diagram. 
Figure \ref{fig_fusion} shows the same data as the diagrams before, plotting 
$G = Q/(Q+5)$ versus the fusion power normalised to the ITER value $2.77(\beta_N/q_{95})^2$.
The dashed lines in this diagram are the different values of the auxiliary heating
power. From left to right 1, 2, 4, and 8 times the ITER design value. 
This diagram shows that high pressure discharges at low $Q$ values would require a large 
amount of installed heating power to be run. It is this diagram that carries the
largest amount of information and we propose it to be used when representing larger 
datasets of advanced scenario discharges. 

\epsfig{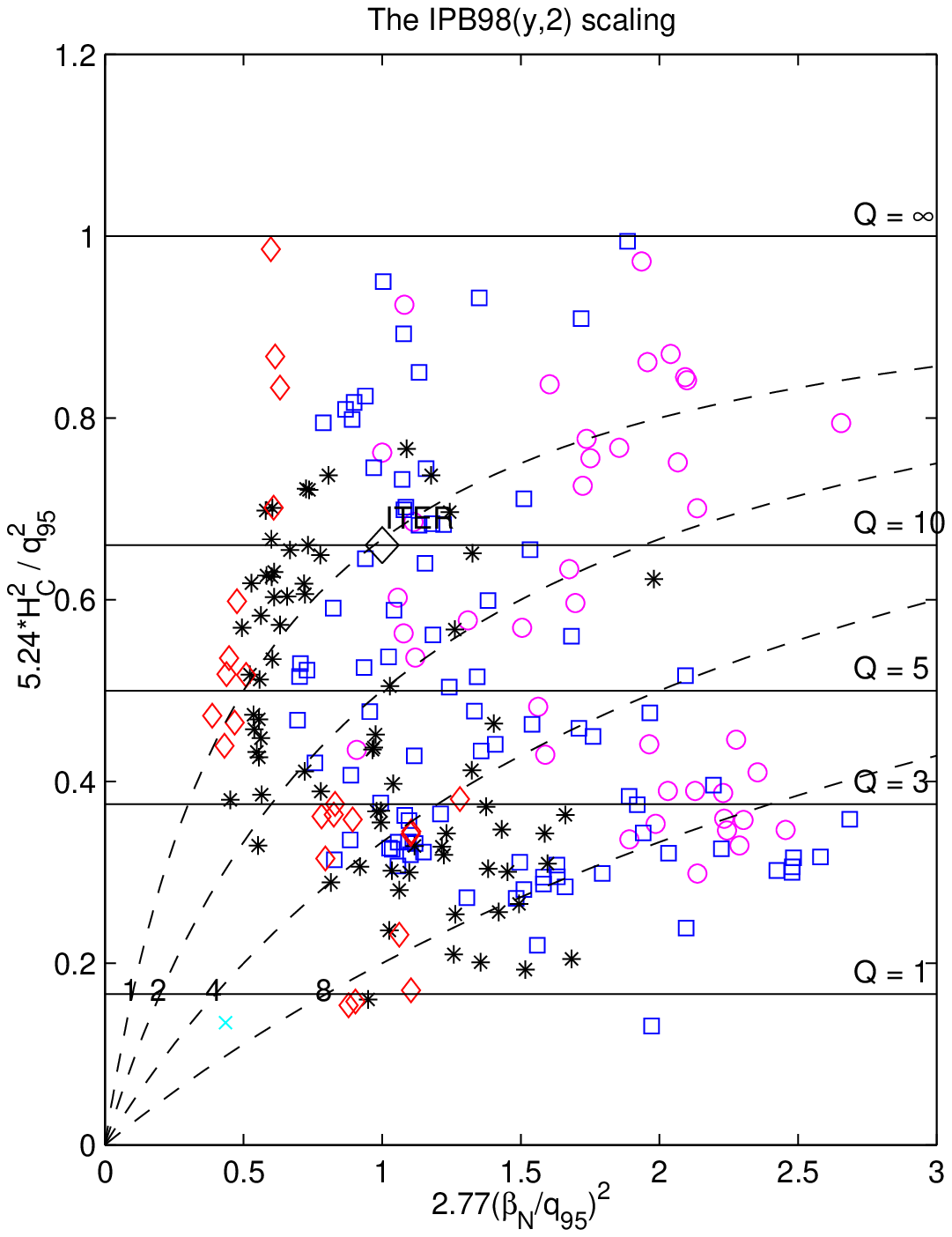}{fig_fusion}{ (color online) Extrapolated values of $Q/(Q+5)$ as a 
function of the normalised fusion power $2.77 (\beta_N/ q_{95})^2$. 
Symbols reflect the $q_{95}$ values as in figure \ref{fig_adv_old}.
The dotted lines indicate the amount of external heating ($P_{aux}$) 
necessary to run the discharge in ITER. From left to right the curves 
correspond to power levels 1, 2, 4, and 8 times the nominal ITER value}

The diagrams presented in this paper are far from perfect, with several effects not 
properly accounted for:
Radiated power due to Bremsstrahlung as well as dilution of the fuel are not properly 
scaled. 
The approach with constant $H$-factor and $\beta_N$ is always daring for an
extrapolation. 
If better MHD stability is, for instance, reached through current profile shaping, then 
one should investigate if such a shaping is extrapolatable to reactor parameters. 
Also, although $\beta_N$ is a good scaling quantity for ideal MHD instabilities,
it does not provide a very good scaling for the NTM, which is often found to limit
the attainable beta. 
Nevertheless for the representation of large data sets these diagrams are certainly 
useful. 
Also they give an idea of what parameters are important when developing
scenarios.

\section{CONCLUSIONS}

In this paper a scaling for tokamak discharges is derived that directly 
measures the fusion gain $G = Q/(Q+5)$, and which are consistent 
with the underlying scaling laws.
It is shown that $\beta_N$ does not have a positive influence on $G$, although
it does of course extrapolate to a larger fusion power. 
Care is to be taken with figures of merit like $H \beta_N / q_{95}^2$. 
Although this figure of merit does measure a combination of good confinement and 
high fusion power, the ITER target value of such a quantity does not automatically
imply a discharge with a sufficiently high $Q$, and might not be attainable with
the limited heating power that will be installed. 
A proposal is made for a graphical representation in which both the extrapolated 
$Q$ as well as the fusion power or the bootstrap fraction can be directly assessed.

%\vfill \eject

\bibliographystyle{iaea}

\end{document}